\documentclass[a4paper]{llncs}

\usepackage{graphicx}
\usepackage{gensymb}
\usepackage{algpseudocode}
\usepackage{multicol}
\usepackage{longtable}

\newenvironment{snippet}
{
\hspace{-\parindent}
\begin{longtable}{p{0.96\textwidth}}
}
{
\hline
\end{longtable}
}

\newcommand{\procedurespace}{\vspace{-10mm}}

\newcommand{\tab}{\hspace*{3mm}}

\title{\LARGE \bf
A Library for Implementing the Multiple Hypothesis Tracking Algorithm
}

\author{David Miguel Antunes\inst{1} \and David Martins de Matos\inst{2} \and Jos\'{e} Gaspar\inst{3}}

\institute{
Institute for Systems and Robotics, $L^2F$, \email{davidmiguel [at] antunes.net}
\and
$L^2F$ - INESC-ID, \email{david.matos [at] l2f.inesc-id.pt}
\and
Institute for Systems and Robotics, IST/UTL, \email{jag [at] isr.ist.utl.pt}}

\begin{document}

\maketitle
\thispagestyle{empty}
\pagestyle{empty}

\begin{abstract}
The Multiple Hypothesis Tracking (MHT) algorithm is known to produce good results in difficult multi-target tracking situations. However, its implementation is not trivial, and is associated with a significant programming effort, code size and long implementation time. We propose a library which addresses these problems by providing a domain independent implementation of the most complex MHT operations. We also address the problem of applying clustering in domain independent manner.
\end{abstract}

\section{Introduction}
The Multiple Hypothesis Tracking (MHT) algorithm, proposed by Reid \cite{Reid79analgorithm} is fundamental in the multi-target tracking field with many applications in different areas. And, even though its significant computational complexity inhibited its utilization for some time, continuous increases in the processing power of computers accompanied by a reduction in their cost, along with some algorithmic improvements \cite{K-Best}, make the implementation of the MHT feasible nowadays \cite{EfficientMHT}. The MHT is now the preferred method for difficult tracking problems \cite{BlackmanTutorial}.

Although the MHT presents advantages when compared to other methods its implementation is not trivial, and requires significant programming effort, extensive code size, and debugging effort \cite{TheProbabilis}.
To address the difficulty in implementing the MHT, we propose and describe a library which can be used to ease this task. Because the MHT is applied in multiple domains, with variations on the algorithm, such as group tracking \cite{TrackingGroup}, the library should not enforce a particular type of MHT implementation. Therefore, it only addresses the management of the multiple hypotheses, and not the probability modeling, movement models, sensor models, and other aspects which may vary across different MHT implementations. 

Cybenko et al. \cite{WhatIsTrackable} identify less usual tracking domains, such as computer security, or social networks. Because the proposed library is not tied to a particular domain, it may be used to apply tracking to domains like the ones identified by these authors.

As noted in the seminal work of Reid \cite{Reid79analgorithm}, an efficient MHT implementation must perform clustering. Clustering consists in dividing the tracking problem into independent sub-problems which can be solved separately. Reid identified the necessary steps to implement clustering in the MHT. However, because the proposed library is independent of the specific domain of the application, and the clustering method described by Reid is specific for a particular domain, a different, more general clustering method is required. We propose such a clustering method for the library.

The work on Process Query Systems (PQS) \cite{ProcessQuerySystemsIntro} by various persons of the Dartmouth College is close in intent with this one. PQS allow the modeling of processes with states, dynamics, and observables in a domain independent manner. The PQS framework is then able to detect and track the evolution of several processes in the environment. However, the PQS framework does not seem to provide clustering, and is tied to the notion of process which may not be appropriate for some multiple hypothesis applications, such as Multiple Hypothesis Situation Analysis \cite{JeanRoyMHSA}.
Jean Roy et al. proposed a domain independent clustering method for the MHT \cite{JeanRoyClusters}, however he does not provide any description of a domain independent MHT library.

The main contribution of this paper is the description of a library which can be used to simplify and speed up the implementation of the MHT independently of the application's domain,  and a method to provide the ability of clustering in such a library.

The proposed library was applied to several problems which are discussed, one of which is detailed in section \ref{sec:implementation}.
Section \ref{sec:overview} gives an overview of the proposed library, section \ref{sec:clustering_method} details the clustering method, section \ref{sec:information_modeling} describes the modeling of information, and the library's operations and data-structures are described in section \ref{sec:library_operations}.

More information and resources on the proposed library are available at \texttt{http://www.multiplehypothesis.com}.

\section{MHT Overview}
\label{sec:overview}

\ifx \ommitimages  \undefined
\begin{figure}
\begin{minipage}[b]{0.30\textwidth}
\centering
\includegraphics[scale=0.8]{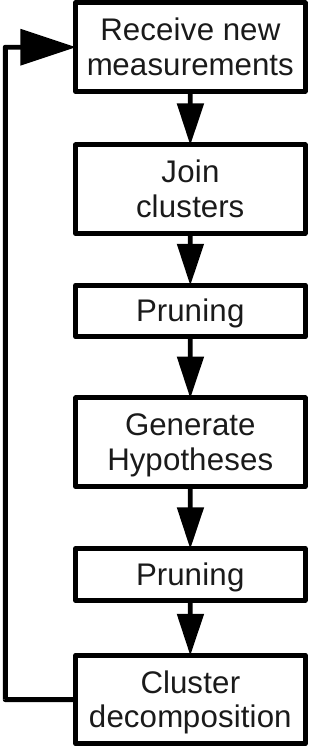}
\vspace{4.65mm}
\caption{Traditional MHT.}
\label{fig:workflow_mht}
\vspace{7.65mm}
\end{minipage}
\hspace{0\textwidth}
\begin{minipage}[b]{0.70\textwidth}
\centering
\includegraphics[scale=0.8]{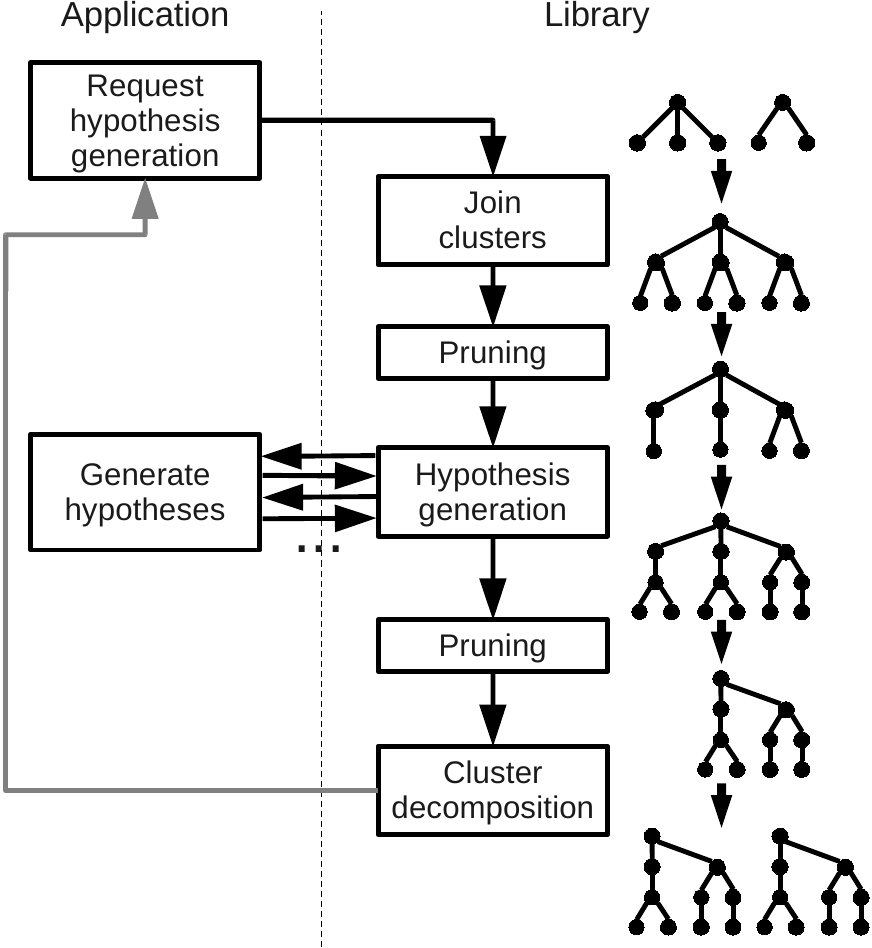}
\caption{Proposed library workflow, and an abstract example of how the operations affect the maintained data structures (right).}
\label{fig:workflow_mhl}
\end{minipage}
\end{figure}
\fi

The usual flow diagram for the MHT is presented in figure \ref{fig:workflow_mht}.
First, the application receives a new set of measurements from the sensor.
Each measurement is either associated with an existing cluster or else a new cluster is created.
When a measurement is associated with more than one cluster they are joined and the resulting cluster is pruned. 
Afterwards, new hypotheses are generated for each set of measurements associated with the same cluster, followed by cluster tree pruning. 
Finally, the clusters may be decomposed according to the rule: if two tracks share a common measurement, they must be on the same cluster.

The operations of the library (shown in figure \ref{fig:workflow_mhl}) are similar to the ones of the traditional MHT.
There is, however, a clear separation of responsibilities between the application and the library.
The main responsibility of the application is the hypothesis generation, as it is domain dependent. The library handles the complicated operations which must modify complex data-structures, freeing the library user from the most difficult and time consuming programming tasks.
How the hypothesis generation is performed in detail, as well as other necessary operations, will be further discussed in the rest of the paper.

\section{Clustering Method}
\label{sec:clustering_method}

Before further detailing the library, the proposed clustering method will be introduced. 

Because the library is independent of the application domain it only \textit{sees} generic information. To ease application development, this information is separated in history and state of the world. It is not necessary for the application to take advantage of this separation, however it has been found useful in structuring the development of tracking applications. For now, consider that the library maintains a set of information $I$, constituted of pieces of information $i_1, i_2, ..., i_{|I|}$. The specific modeling of information and its separation in history and state of the world, will be discussed later, in section \ref{sec:information_modeling}.

Clustering on the library is based on two rules. The first one is the following:
\textit{If $i_1$ and $i_2$ where generated at the same hypothesis generation, then they must remain in the same cluster.}
And the second rule is:
\textit{If, to assert $i_1$, one needs to know $i_2$, then both must remain in the same cluster.}
The algorithm operations which concern clustering try to form the maximum number of clusters, while still respecting these two rules.

The validity of the proposed clustering rules will now be discussed.

\ifx \ommitimages  \undefined
\begin{figure}
\begin{minipage}[b]{0.6\textwidth}
    \centering
    \includegraphics[scale=0.8]{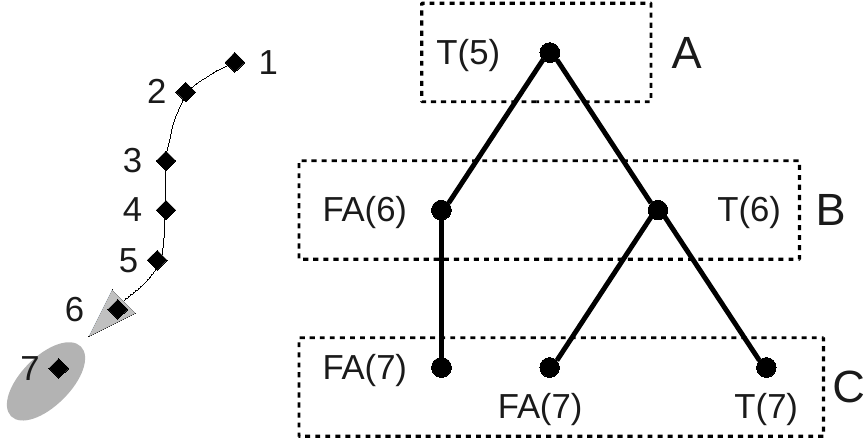}
    \caption{Tracking scenario (left), and an abstract representation of a possible hypothesis tree (right).}
    \label{fig:single_track}
\end{minipage}
\hspace{0\textwidth}
\begin{minipage}[b]{0.4\textwidth}
	\centering
	\includegraphics[scale=0.8]{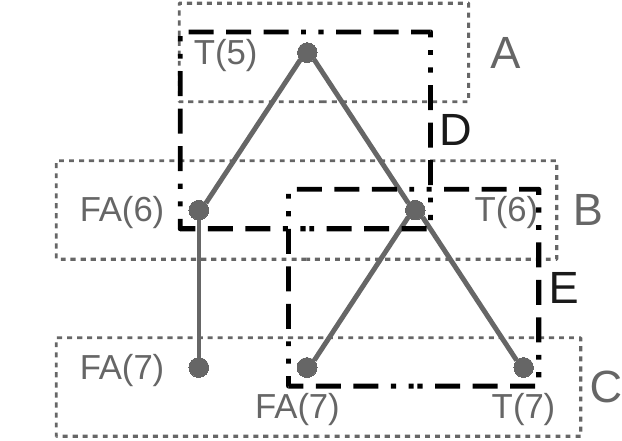}
	\caption{Applying the second clustering rule.}
	\label{fig:second_rule2}
\end{minipage}
%\hspace{0.5cm}
\end{figure}
\fi

\subsection{Validity of the First Rule} In the MHT, in an hypothesis generation, no more than one hypothesis can be correct, and exactly one of the hypotheses will be considered correct, i.e. only one will survive pruning in the long run. Consider the scenario in figure \ref{fig:single_track}, where target \texttt{T} is being tracked. Measurements are denoted by numbers 1 to 7. At the right there is an abstract representation of a possible hypothesis tree (or cluster). The first rule dictates that the information inside \texttt{A}, \texttt{B}, and \texttt{C}, cannot be separated to different clusters. A separation of this information could lead to inconsistent situations.
For example, consider \texttt{FA(6)} (meaning measurement 6 is a false alarm) and \texttt{T(6)} (meaning measurement 6 is a detection of target \texttt{T}) on \texttt{B} are separated to different clusters.
Then, there would be a global\footnote{
Every cluster contains information only on a specific part of the world. A global hypothesis contains information on all of the world, and it is built by selecting an hypothesis from every cluster \cite{BlackmanTutorial}.	
} hypothesis where both would be true that is, a measurement could be both a false alarm and a detection of target \texttt{T} at the same time. 
And there would also be a global hypothesis where none would be present, resulting in any information related to measurement 6 disappearing, if only this hypothesis survived pruning.
Thus, the rule first rule ensures that no more and no less than one of the pieces of information generated in one hypothesis generation will be present in each possible global hypothesis.

From now on, the sets of information which must be on the same cluster, such as \texttt{A}, \texttt{B}, and \texttt{C} will be denoted as constraints.

\subsection{Validity of the Second Rule} 
To apply the second rule, the library needs to know which information the application needs to generate a certain hypothesis. In practice, this is implemented in the following manner. Before generating new hypotheses, the application identifies the information which will be required to generate those hypotheses (a subset of all the information the library is storing) and requests it from the library. When the application generates new hypotheses for a leaf in a cluster's hypothesis tree, the library provides it with the subset of the required information which is true in that leaf. According to the second rule, the information provided by the library and the information generated by the application must be kept on the same cluster.

The constraints introduced by the first rule alone are not sufficient to correctly split clusters. Consider the example in figure \ref{fig:single_track}. Although the information in constraints \texttt{A}, \texttt{B}, and \texttt{C} cannot be separated, the hole constraints may be separated to different clusters. In which case, for example, the following global hypothesis would be possible: \{ \texttt{T(5)}, \texttt{FA(6)}, \texttt{T(7)} \}. This hypothesis is incorrect as \texttt{T(7)} only makes sense with \texttt{T(6)}. With only the first rule it is possible that in a global hypothesis there are two pieces of information incompatible with each other, or that there is a missing piece of information (such as \texttt{T(6)}). However, when generating hypotheses, the application must (and needs to) request any piece of information which would prevent the generated hypotheses from making sense, or any piece of information which it requires to know to generate the hypotheses (such as \texttt{T(6)}). Then, the constraints introduced by the second rule ensure that there are no incorrect global hypotheses. In the example of figure \ref{fig:single_track}, the application would request the information on targets near measurement 7 (corresponding to \texttt{T(6)}) and, applying the second rule, constraint \texttt{E} would be introduced. The case for constraint \texttt{D} is similar.

\section{Information Modeling}
\label{sec:information_modeling}

As the library is generic and it is intended to be applicable in multiple domains the modeling of the information contained in the library is also generic. The library can store and manage two different types of information: events and facts. The library does not contain any predefined events nor facts, they are defined by the application in order to fit its needs.

\subsection{Events}

An event corresponds to a piece of history of the world, and contains information about what (possibly) happened in the world at some point in time. For example ``ship 9372 moved to (1232.2, 293.6) at 15:53AM'', or ``sensor with id=173 produced a false alarm at 04:26PM''.

\subsection{Facts}

In order to request the necessary information from the library, prior to an hypothesis generation, the application has to identify and find the events which will be relevant for that hypothesis generation. For example, following a detection at $(x,y)$ by a ship's radar, it is necessary to identity the events which place a ship near $(x,y)$, as it might be the cause of that detection. And it may happen that a ship was detected in that area two minutes, half an hour, or even three hours ago. Thus, the application may need to search trough many events to find the ones it requires. The same happens if the application has to answer the question: ``where is ship 9372?'' or ``how many ships are in the $(x_0, y_0, x_1, y_1)$ area?''.

To simplify the answer to these questions, the application may define facts. Whereas events are part of the history of the world, facts represent the current state of the world. For example, when the application generates the event ``ship 9372 moved to (1232.2, 293.6) at 15:53AM'', it may also generate the fact ``ship 9372 is in (1232.2, 293.6)''. Contrary to events, facts do not usually include a timestamp, as a fact is true as long as it exists. And, while events are never deleted (unless they are pruned), facts have a limited lifetime. In the example, when the ship moves a new fact should be created with the new position, and the old one deleted. In the example, if there is a fact per ship, answering the aforementioned questions becomes simpler as there is one, and only one, fact for each ship which contains the most up-to-date information on the ship position, velocity, etc.

\section{The Library's Operations}
\label{sec:library_operations}

The library's main operations which will now be detailed are depicted in figure \ref{fig:workflow_mhl}, and they are: cluster joining, hypothesis generation, cluster splitting, and pruning. However, before detailing these operations, the data structures upon which they operate will be described.

\subsection{Data Structures}
\label{sec:operations_detail}

\ifx \ommitimages  \undefined
\begin{figure}[thpb]
    \centering
    \includegraphics[scale=0.8]{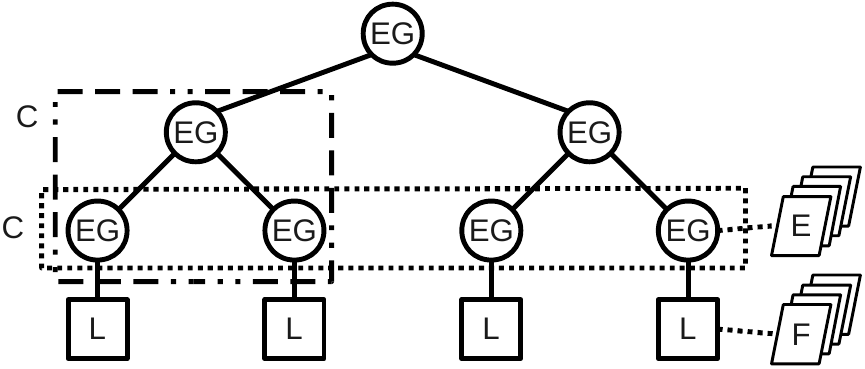}
    \caption{Example of the tree structure (or cluster).}
    \label{fig:tree_structure}
\end{figure}
\fi

The fundamental data structure in the library is the hypothesis tree (or cluster). An example tree is depicted in figure \ref{fig:tree_structure}. 
The events in a generated hypothesis can be treated in the same manner for most of the operations, so, they are aggregated in event groups.
The nodes in the hypothesis tree are constituted of event groups (denoted by EG in figure \ref{fig:tree_structure}), containing events (denoted by E). The tree also contains constraints (denoted by C) and, associated with each event group at the bottom of the tree, there is a structure denoted by leaf (denoted by L). Facts (denoted by F) are stored in the leaves.
The library maintains several of these trees at the same time.

Additionally there is a relation between each fact and an event group, not shown in the figure. For each generated hypothesis, its events are stored in a new event group and the facts added to a new leaf. Then, each of the facts is associated with the new event group. We say the fact ``depends upon'' the event group. This association dictates, for example, the fact's probability, which is the same that the probability of its associated event group.

In section \ref{sec:clustering_method}, a simplification was made by considering that the library only keeps ``information''. Now the clustering rules will be stated directly in terms of events, facts, event groups, and constraints. The first rule can be formulated as: \textit{In an hypothesis generation, a new constraint must be created containing all the new event groups.} And the second rule corresponds to: \textit{When generating hypotheses for a leaf, a new constraint must be created with: the new event groups; the leaf's event groups containing requested events; and the leaf's event groups to which requested facts are associated.} The event groups of a leaf are all the event groups which are in the path from the leaf up to the root of the hypothesis tree.

\subsection{Operations}
An overview of the operations depicted in figure \ref{fig:workflow_mhl}, and which will be detailed next is as follows. After receiving a new measurement, the application needs to generate hypotheses on the origin and implications of the new measurement. The application knows all the facts and events contained in the library, and selects the ones which are relevant for the hypothesis generation. The library receives the relevant events and facts for that hypothesis generation and joins all the clusters which contain them into one single cluster, then, requests the application to generate new hypotheses for each leaf of the cluster, providing the application with the subset of the requested events and facts which are true according to that leaf. Afterwards the cluster is pruned and, if it can now be divided into clusters containing subsets of the events, facts, and constraints in the original one, while still complying to the clustering rules, then splitting is applied.

Procedure \ref{sec:operations_detail}.1, \textsc{JoinClusters}, details the method for joining different clusters. Pruning is applied each time two clusters are joined (the representation in figure \ref{fig:workflow_mhl} simplifies this by showing pruning only once, after cluster joining).
The procedure \textsc{UnifyConstraints} is called by the previous one to guarantee that, for every constraint $c$ in cluster $c2$ (see procedure \ref{sec:operations_detail}.1) containing the set of events $e$, a constraint $C$ exists in the final cluster which contains the set $E$ of all the events of the clones of constraint $c$. Thus, $|E| = |e| \times |c1.leaves|$. If no events or facts are requested, a new cluster is created (cluster joining is not necessary).

The procedure \ref{sec:operations_detail}.3, \textsc{HypGen}, asks the application to generate hypotheses for each leaf of the cluster. When generating hypotheses for a leaf, it provides the application with the subset of the requested facts and events which are available in the leaf. The generation will result in the tree being augmented with new event groups and new leaves. The \textsc{HypGen} procedure also creates the appropriate new constraints and prunes the resulting tree.

Several pruning strategies may be applied. The pruning operation can limit the number of leaves, the depth of the tree, or follow other application specific strategy. When a leaf is pruned, procedure \ref{sec:operations_detail}.4, \textsc{RemoveLeaf}, is called to remove it from the cluster. When a brach of the tree is pruned, \textsc{RemoveLeaf} is called for every leaf coming from that branch.

Cluster splitting can be performed if non-intersecting sets of constraints are found. If constraints $c1$ and $c2$ contain a common event group they intersect. For each set of independent constraints, $C$, a new cluster, $nCluster$, can be created by cloning the original cluster. In $nCluster$, the events in event groups which are not contained in any $c \in C$ are deleted. Also, only facts associated with event groups in $c \in C$ are kept, and $nCluster$ only contains the constraints in $C$. Procedure \ref{sec:operations_detail}.5, \textsc{Split}, details this operation.

\vspace{-5mm}
\begin{snippet}
\textbf{Procedure 1} Join clusters with required events or facts\\  \hline

\textbf{procedure} \textsc{JoinClusters($reqEvents$, $reqFacts$)}\\
Find the clusters, $clusters$, containing required events or facts;\\
Select a cluster $c1 \in clusters$;\\
For each $c2 \neq c1$ in $clusters$:\\
\tab $clones \leftarrow \{\}$; $newLeaves \leftarrow \{\}$;\\
\tab For each $leaf$ in $c1.leaves$:\\
\tab \tab $clone \leftarrow$ clone of $c2$; Add $clone$ to $clones$;\\
\tab \tab Add $c2.rootEventGroup$ to $leaf.eventGroup$ children;\\
\tab \tab For each $cLeaf$ in $clone.leaves$:\\
\tab \tab \tab Add $cLeaf$ to $newLeaves$; Add $leaf.facts$ to $cLeaf.facts$;\\
\tab \tab \tab $cLeaf.prob \leftarrow cLeaf.prob \times leaf.prob$;\\
\tab Delete $c2$; $c1.leaves \leftarrow newLeaves$; \textsc{UnifyConstraints} ($c1$, $clones$); Prune $c1$;\\
Return $c1$;\\
\end{snippet}

\procedurespace
\begin{snippet}
\textbf{Procedure 2} Connect constraints in cloned clusters\\ \hline

\textbf{procedure} \textsc{UnifyConstraints($c1$, $clones$)}\\
For $i \leftarrow 1$ up to $|clones[1].constraints|$:\\
\tab Create constraint $newC$;\\
\tab For each $clone$ in $clones$:\\
\tab \tab Add all event groups in $clone.constraints[i]$ to $newC.eventGroups$;\\
\tab Add $newC$ to $c1.constraints$;\\
\end{snippet}

\procedurespace
\begin{snippet}
\textbf{Procedure 3} Generate hypotheses\\ \hline

\textbf{procedure} \textsc{HypGen}($cluster$, $reqEvents$, $reqFacts$)\\
$newEventGroups \leftarrow \{\}$\\
For each $leaf$ in $cluster$:\\
\tab $provFacts \leftarrow leaf.facts \cap reqFacts$; $provEvents \leftarrow leaf.events \cap reqEvents$;\\
\tab Create a new constraint, $cstrnt$, with the event groups containing provided events, and the event groups to which provided facts are associated;\\
\tab Request the application to generate hypotheses, providing it with $provEvents$ and $provFacts$;\\
\tab For each generated hypothesis, $hyp$:\\
\tab \tab Create a new event group, $nEventGroup$, with $hyp.events$;\\
\tab \tab Create a new leaf, $nLeaf$, with $hyp.facts \cup (leaf.facts\setminus provFacts)$;\\
\tab \tab Add $nLeaf$ to $cluster.leaves$;\\
\tab \tab Add $nEventGroup$ to $leaf.eventGroup$ children, to $newEventGroups$, and to $cstrnt$;\\
\tab \tab Set $nLeaf.probability$ to $hyp.probability * leaf.probability$;\\
\tab Remove $leaf$ from $cluster$;\\
Create a new constraint with $newEventGroups$; Normalize leaves probability; Prune $cluster$;\\
\end{snippet}

\procedurespace
\begin{snippet}
\textbf{Procedure 4} Remove leaf from cluster\\ \hline

\textbf{procedure} \textsc{RemoveLeaf}($cluster$, $leaf$)\\
Remove $leaf$ from $cluster.leaves$; $parent \leftarrow leaf.eventGroup$;\\
While $|parent.children| = 0$ do:\\
\tab Remove $parent$ from all $cluster.constraints$, and from $parent.parent.children$;\\
\tab $parent \leftarrow parent.parent$;\\
Remove empty constraints in $cluster.constraints$; Normalize leaves probabilities;\\
\end{snippet}

\procedurespace
\begin{snippet}
\textbf{Procedure 5} Cluster splitting\\ \hline

\textbf{procedure} \textsc{Split}($cluster$)\\
$superConsts \leftarrow \{\}$; Let $map$ be a map from constraints to sets of constraints;\\
For each $c$ in $cluster.constraints$:\\
\tab Let $sC$ be a clone of $c$; Add $sC$ to $superConsts$; $map[sC]$ $\leftarrow$ $\{$ $c$ $\}$;\\
For all $sC_1, sC_2 \in superConsts$ $:$ $(sC_1.eventGroups$ $\cap$ $sC_2.eventGroups)$ $\neq$ $\{\}$:\\
\tab Create new constraint, $newConstr$; Add $newConstr$ to $superConsts$.\\
\tab $newConstr.eventGroups \leftarrow$ $sC_1.eventGroups$ $\cup$ $sC_2.eventGroups$;\\
\tab Remove $sC_1$ and $sC_2$ from $superConsts$;\\
\tab $map[newConstr]$ $\leftarrow$ $map[sC_1]$ $\cup$ $map[sC_2]$; Remove $sC_1$ and $sC_2$ from $map$;\\
For each $constraints$ in $map$ keys:\\
\tab Create a new cluster by cloning $cluster$. Keep only constraints $constraints$, the events contained in event groups in $constraints$, and the facts associated with those event groups;\\
Remove $cluster$;\\
\end{snippet}

\subsection{Further Notes}
In order to request events and facts from the library for an hypothesis generation, the application may obtain all the events and facts the library contains and search for the ones of interest. However, a better approach is for the library to notify the application about changes to its available events and facts. Then the application can organize them in a way which speeds up searches. For example, the application may organize facts containing the position of ships so that it is efficient to find all the ships in a certain area.

The library does not keep the events which have a probability of 1. These are the events at the root of the clusters for which there is no uncertainty. Therefore, after a cluster is pruned, they should be sent to the application and removed from the cluster. Then, the application decides what to do with them (storing them in persistent storage, for example).

Facts can be seen as a special events. They have only two differences: contrary to events, facts are deleted after they are provided to the application in an hypothesis generation (but they may be generated again); when the event group associated with a fact reaches the root of the tree, with probability of 1, it is not sent to the application, instead, a new cluster is created with only that fact.

It may happen that, due to cluster splitting, a cluster is created containing only facts and/or events which will never be requested by application. For example a cluster which only contains false alarm events. Because it will never be requested it is wasting memory space. Thus the application should provide a predicate function which accepts the facts and events of a cluster and asserts if that cluster can ever be required for hypothesis generation. If not, then best K hypothesis pruning is applied to the cluster, with K$=1$.

\section{Radar Tracking Application}
\label{sec:implementation}

The proposed library was implemented in Java and applied to several tracking problems. One of the library applications is a radar tracker simulator, developed to illustrate the capabilities of the library in a self-contained manner, so that other members of the research community may become rapidly familiarized with the library work-flow by analyzing the application code. This simulator will now be described in detail.

\ifx \ommitimages  \undefined
\begin{figure}
\begin{minipage}[b]{0.325\textwidth}
\centering
\includegraphics[width=3.5cm]{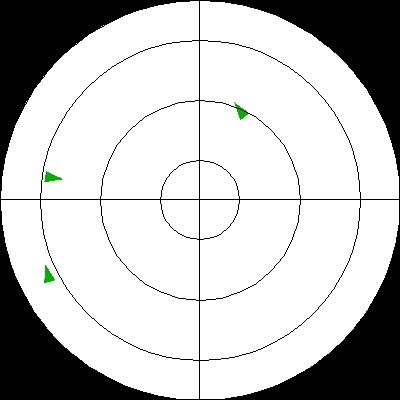}
\caption{Ground truth.}
\label{fig:radar-gt}
\end{minipage}
\begin{minipage}[b]{0.325\textwidth}
\centering
\includegraphics[width=3.5cm]{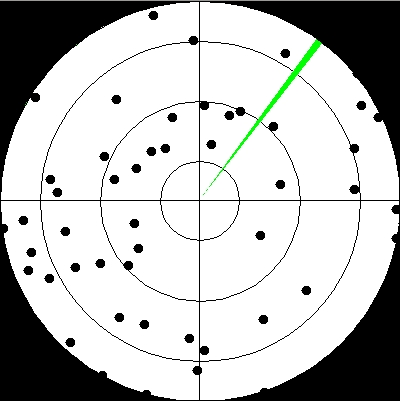}
\caption{Radar.}
\label{fig:radar-radar}
\end{minipage}
\begin{minipage}[b]{0.325\textwidth}
\centering
\includegraphics[width=3.5cm]{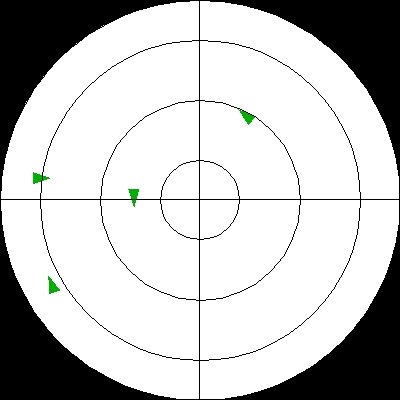}
\caption{Tracker output.}
\label{fig:radar-tracker}
\end{minipage}
\end{figure}
\fi

The simulated radar is stationary and detects ships in a circular region around it. 
Every time the radar beam ends a full 360\degree turn the radar detections are provided to the tracker.
A Kalman Filter with a constant velocity model is used to predict the position of the targets, but the simulated targets have random velocity and direction changes.

The problem was modeled with facts and events. Only one fact (named \textit{TargetPositionFact}) was required to model the problem. This fact contains the target identifier number, the state of the target's Kalman Filter, and the time of the target's last detection (which is used to terminate the track if the target is not detected after a certain period of time).

To keep the tracking history four events were defined. 
A \textit{TrackInitiatedEvent} containing the position and the identifier of the target which initiated the track. 
A \textit{TrackTerminatedEvent} containing the identifier of the target whose track terminated. 
A \textit{TargetMovedEvent} containing the initial and final positions of the target, and the target identifier. 
And a \textit{FalseAlarmEvent} with the position of the detection which was considered a false alarm. All these events contain a timestamp.

After the tracker receives a new set of measurements, when the radar completes a full 360\degree turn, it needs to generate hypotheses. Only facts are required from the library for hypothesis generation. For this reason, the tracker can function with or without events. Without events it does not keep history, but is faster.
When generating hypotheses for a measurement if a target, represented by a \textit{TargetPositionFact}, contains the measurement in its validation gate then the fact must be required (the measurement may be a detection of that target). Furthermore, if the same \textit{TargetPositionFact} must be required for two measurements, their hypothesis generation must be made at the same time. Each measurement either is a false alarm, a new target detection, or a detection of an existing target.

Due to clustering the tracker is able to track many targets at the same time, while still taking an acceptable time to process each set of detections coming from the radar. Figure \ref{fig:timeToProcess} shows the time to process the detections from the radar with different number of targets, it grows almost in a linear fashion.

\ifx \ommitimages  \undefined
\begin{figure}
\begin{minipage}[b]{1\textwidth}
\centering
\includegraphics[width=0.45\textwidth]{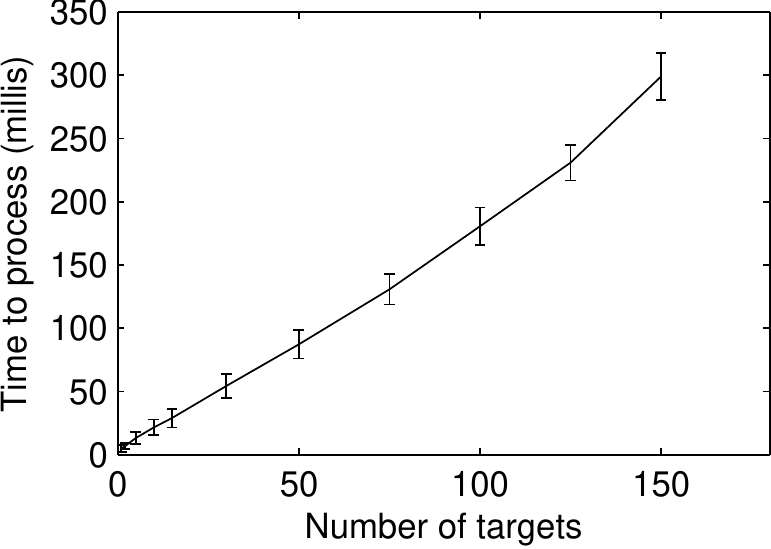}
\caption{Time to process new detections. Error bars show standard deviation.}
\label{fig:timeToProcess}
\end{minipage}
\end{figure}
\fi

The proposed library was implemented in Java and made available on-line together with the demonstration of the radar tracking application.

\section{Conclusions}
\label{sec:conclusions}

In this paper we address the problem of the complexity of implementing the MHT by proposing a library which implements much of the algorithm, leaving only the domain specific tasks to the application developer. And, even though the library is independent of the application domain, we provide a method for applying clustering, also domain independent, which is essential for an efficient implementation of the MHT. Additionally, the library provides an explicit separation of the state and history of the world, which can be very useful in structuring tracking applications.

In order to demonstrate the effectiveness of the library a radar tracking application was developed using the library. Simulations have shown successful tracking of the targets with a computation time growing in an approximately linear manner with the number of targets, due to the clustering ability of the library.

\bibliographystyle{splncs}
\bibliography{GMH}

\end{document}